\def \cm{~\rm{cm}}
\def \s{~\rm{s}}
\def \km{~\rm{km}}

\def \K{~\rm{K}}
\def \g{~\rm{g}}

\def \AU{~\rm{AU}}

\def \yr{~\rm{yr}}

\documentclass[12pt,preprint]{aastex}
\usepackage{natbib}
\usepackage{epsfig}

\shorttitle{Binay Accretion}
\shortauthors{Soker}
\slugcomment{Draft version of \today}

\begin{document}

\title{Wind Accretion by a Binary Stellar System and Disk Formation}

\author{Noam Soker \altaffilmark{}}
\altaffiltext{1}
{Department of Physics, Technion$-$Israel Institute of Technology,
Haifa 32000 Israel, and Department of Physics, Oranim, Israel;
soker@physics.technion.ac.il.}

\begin{abstract}

I calculate the specific angular momentum of mass accreted by a binary
system embedded in the dense wind of a mass losing
asymptotic giant branch star.
The accretion flow is of the Bondi-Hoyle-Lyttleton type.
For most of the relevant parameters space the flow is basically
an isothermal high Mach number accretion flow.
I find that when the orbital plane of the accreting binary system
and the orbital plane of the triple system are not parallel
to each other, the accreted mass onto one or two of the
binary system components has high specific angular momentum.
For a large fraction of triple-star systems, accretion disks will
be formed around one or two of the stars in the binary system,
provided that the mass ratio of the two stars in the
accreting binary system is $\gtrsim 0.5$.
Such disks may blow jets which shape the descendant
planetary nebula.
The jets' axis will be almost parallel to the orbital plane
of the triple-star system.
One jet is blown outward relative to the wind, while
the other jet pass near the mass losing star, and is more
likely to be slowed down or deflected.
I find that during the final asymptotic giant branch phase,
when mass loss rate is very high,
accretion disk may form for orbital separation between
the accreting binary systems and the mass losing star of
up to $\sim 400-800 \AU$.
I discuss the implications for the shape of the descendant
planetary nebula, and list several planetary nebulae which
may have been shaped by an accreting binary star system,
i.e., by a triple star system.
\end{abstract}

{\it Subject headings:}
planetary nebulae: general
$-$ stars: binaries
$-$ stars: AGB and post-AGB
$-$ stars: mass loss
$-$ ISM: general


\section{INTRODUCTION}

 Both theory (Soker 1990) and observations (Sahai \& Trauger 1998)
suggest that many, but not all, planetary nebulae (PNs) are shaped by
two oppositelly ejected jets from the progenitor system
(for a recent review see Soker 2004).
If not well collimated, these jets are termed CFW, for collimated
fast wind.
In principle, the jets (or a CFW) can be blown by the post-asymptotic
giant branch (AGB) progenitor itself, or by an accreting companion.
Theoretical and observational considerations, e.g., the detection
of collimated outflows emanating from AGB stars vicinities
(Imai et al.\ 2002, 2003; Hirano et al.\ 2004; Sahai et al.\ 2003;
Vinkovic et al.\ 2004), suggest that in most cases, or even all,
a binary companion blows these jets (see Soker 2004).
The jets, as with most other astrophysical objects, are thought to
be blown when an accretion disk is formed around a compact
object, and the accretion rate is high enough.

For an accretion disk to form, the specific angular momentum of
the accreted mass $j_a$, must obey the condition $j_a > j_b$, where
$j_b=(G M_b R_b)^{1/2}$ is the specific angular momentum of a
particle in a Keplerian orbit at the equator of the accreting
star of radius $R_b$ and mass $M_b$.
The accretion flow is of the Bondi-Hoyle-Lyttleton type
(Hoyle \& Lyttleton 1939; Bondi \& Hoyle 1944),
where gas with an impact parameter
$b \lesssim R_{\rm acc}=2 G M_b /v_r^2$ is accreted.
The impact parameter is the distance of the accreted mass
from the symmetry line of the flow, termed the accretion line,
at infinity before reaching the shock wave,
\begin{eqnarray}
R_{\rm acc} = 10.6 \AU
\left( \frac {M_b}{0.6 M_\odot} \right)
\left( \frac {v_r}{10 \km \s^{-1}} \right)^{-2} ,
\end{eqnarray}
is the Bond-Hoyle accretion radius, and $v_r$ is the relative velocity
between the gas and the accreting body.
For the rest of the paper I will be dealing mostly with large
orbital separations to the mass losing star $a_0$,
such that the orbital velocity is low, and the relative velocity can
be taken as the wind velocity $v_r \simeq v_s$.
The condition for a companion in a circular orbit accreting from the
AGB wind (but not via Roche lobe overflow [RLOF]), can be written in
the following form (Soker 2001)
\begin{eqnarray}
1< \frac {j_a}{j_b} = 0.25
\left( \frac {\eta}{0.2} \right)
\left( \frac {M_a+M_b}{1.2 M_\odot} \right)^{1/2}
\left( \frac {M_b}{0.6 M_\odot} \right)^{3/2}
\left( \frac {R_b}{1 R_\odot} \right)^{-1/2}
\left( \frac {a_0}{100 \AU} \right)^{-3/2}
\left( \frac {v_s}{10 \km \s^{-1}} \right)^{-4} ,
\end{eqnarray}
where $M_a$ is the mass of the mass-losing star.
Here $\eta$ is a parameter indicating the reduction in the
specific angular momentum of the accreted gas because of the
increase in the cross section for accretion from the low-density side.
Namely, gas parcels with larger impact parameter are accreted
from the low-density side.
Livio et al.\ (1986; see also Ruffert 1999) find that
for high Mach number flows  $\eta \sim 0.1$ and $\eta \sim 0.3$,
for isothermal and adiabatic flows, respectively.
The scaling of the masses are for a star about to leave the AGB,
most relevant stars do it with a mass of $M_a \simeq 0.6 M_\odot$,
and taking a companion of equal mass at that point.
For the mass loss rate and velocity of the AGB wind, a short period
of very high mass loss rate and low velocity is assumed
(sometimes referred to as the superwind or final intensive wind [FIW]).

Another plausible condition for the formation of jets
(or a CFW) is that the accretion rate should be above a certain
limit $\dot M_{\rm crit}$,
which I take as $10^{-7} M_\odot \yr^{-1}$ for accretion onto a
main-sequence star and $10^{-8} M_\odot \yr^{-1}$ for accretion
onto a WD (Soker \& Rappaport 2000).
The Bondi-Hoyle-Lyttleton accretion rate is
$\dot M_b \simeq \pi R_{\rm acc} ^2 v_r \rho$,
where the density at the location of the accretor is
$\rho = \vert \dot M_a \vert / (4 \pi a_0^2 v_s)$.
Substituting the relevant parameters during the final
intensive wind, and taking $v_r \simeq v_s$, the mass accretion rate is
\begin{eqnarray}
\dot M_b \simeq
3 \times 10^{-7}
\left( \frac {M_b}{0.6 M_\odot} \right)^{2}
\left( \frac {v_s}{10 \km \s^{-1}} \right)^{-4}
\left( \frac {a_0}{100 \AU} \right)^{-2}
\left( \frac {\vert \dot M_a \vert }{10^{-4} M_\odot \yr^{-1}} \right)
M_\odot \yr^{-1}.
\end{eqnarray}
It is commonly assumed that $\sim 10 \%$ of the accreted mass is
blown into the CFW (or jets; Soker \& Rappaport 2000), having speeds
of the order of the escape velocity from the companion
(Livio 2000), $100-10^3 \km \s^{-1}$ and $10^3-10^4 \km \s^{-1}$,
for a main-sequence and a WD companion, respectively.

The basics of the binary model for shaping PNs and related
objects, e.g., $\eta$ Carinae, are reviewed in
Soker (2004, where more references are given), while the
ejection of two jets by a wide binary companion, which is one
of several routes for the shaping of PNs and which is a process
directly relevant to the present paper, is studied in Soker
(2001; see also Soker \& Rappaport 2000).
Here I notice the following.
(1) By equation (2), accreting WD companions may form
accretion disks at orbital separation of $a_0 \lesssim 150 \AU$,
while main sequence stars must be much closer at $a_0 \lesssim 40 \AU$.
(2) When the mass losing upper-AGB star lose mass at a very
high rate of $\dot M \sim 10^{\-4} M_\odot \yr^{-1}$, the condition
on the angular momentum (equation 2) is more difficult to fulfill
than the condition on the mass accretion rate.
(3) A single companion star will blow jets perpendicular to
the binary equatorial plane.
(or the jets will precess around this direction).
In the following sections I'll show that these three properties
not necessarily hold for a binary system accreting from the wind
of AGB stars.
Readers interested in results and observational implications only,
may skip to section 4.

\section{THE FLOW STRUCTURE}

The basic structure of the triple-star system is drawn
schematically in Figure 1, where I refer to stars by their masses.
The binary system $M_{b1}-M_{b2}$ is accreting mass from the wind
blown by the star $M_a$.
The orbital separation between $M_a$ and the center of mass of the
accreting binary system is $a_0$, while $a_{12}$ is the orbital separation
of the accreting binary system. I assume circular orbits, with
an inclination angle $\theta$ between the two orbital planes defined here.
\begin{figure}
\plotone{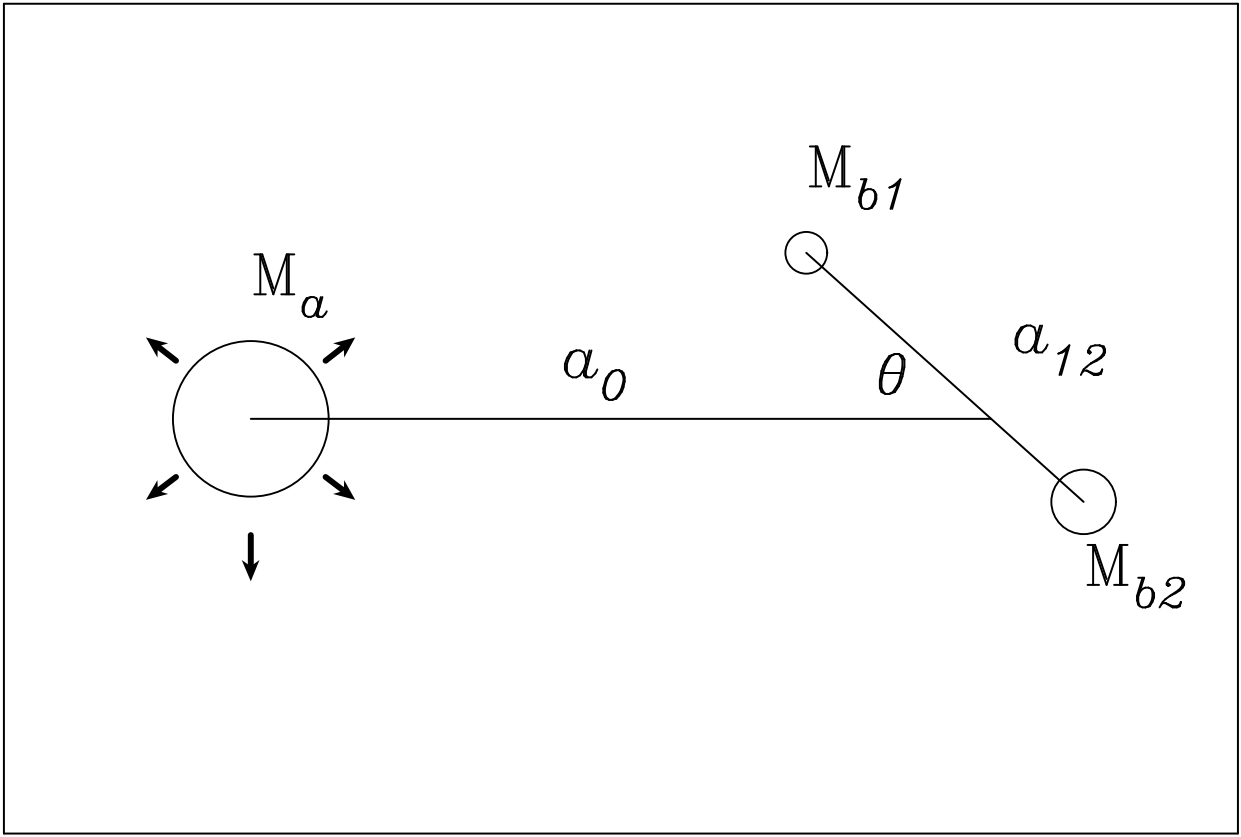}
\vskip 0.2 cm
\caption{A schematic drawing of the triple star system.
The binary system $M_{b1}-M_{b2}$ accretes mass from the wind
blown by the star $M_a$.
Circular orbits are assumed, with an inclination angle $\theta$
between the two orbital planes defined here.  }
\end{figure}

The Bondi-Hoyle-Lyttleton accretion flow structure onto a single star was
calculated analytically and numerically, in
two and three dimensions, in many papers.
A recent relevant paper which contains many references to earlier
papers is Pogorelov, Ohsugi, \& Matsuda (2000; see also
Foglizzo \& Ruffert 1999).
The schematic flow structure is drawn in Figure 2,
which presents the flow in a plane containing the accretion line.
The arrows present the flow of gas around the gravitating system.
The unperturbed gas flows from left to right, and hits a shock
wave as drawn.
Gas with impact parameter smaller than a value of
$b_c \simeq R_{\rm acc}$
is accreted; in the figure these are streamlines with impact
parameter smaller than that of the streamlines represented
by 5 consecutive arrows.
Behind the shock wave gas changes direction, and either
flows to infinity, or flows toward the
accreting system directly, or first to the accretion line
behind the accreting system, and then toward the accreting body.
A high density flow is formed behind the accreting system
along and near the accretion line (represented by 3 close arrows).
This is termed the accretion column.
The exact structure of the flow and the accretion rate depend
on the adiabatic index $1\leq \gamma \leq 5/3$
(Pogorelov et al.\ 2000; for an isothermal flow $\gamma=1$;
for adiabatic flow $\gamma=5/3$).
The flow structure is characterized by, among other things,
the structure of the shock wave (small or wide opening),
the flow behind the shock, and the density and velocity
in the accretion column.
For lower values of $\gamma$ the shock wave is narrower and
the accretion column denser.
The accretion column becomes narrower for higher Mach numbers
(Pogorelov et al.\ 2000).
The gas temperature at large distances $r$ from the stellar surface
of AGB stars is given by (e.g., Frank 1995)
$T(r) \simeq 670 (T_a/3,000 \K)(r/10 R_a)^{-1/2} \K$,
where $R_a$ is the radius of the mass losing star, and $T_a$ is
its surface temperature.
The corresponding sound speed is
$C \simeq 3 (T_a/3,000 \K)^{-1/2} (r/10 R_a)^{1/4} \km \s^{-1}$.
The wind Mach number for most relevant cases is Mach$\gtrsim 5$.
\begin{figure}
\plotone{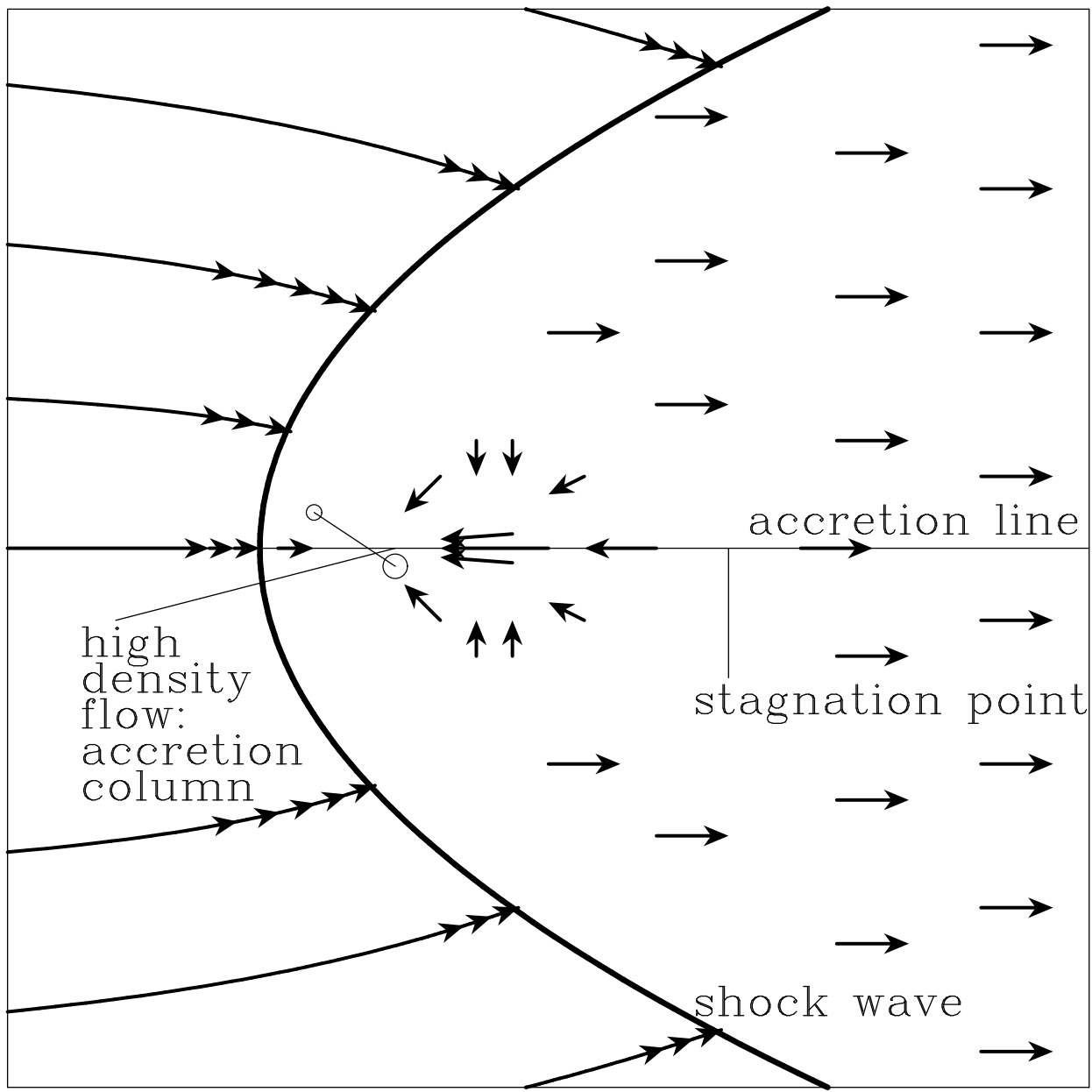}
\vskip 0.2 cm
\caption{A schematic drawing of the flow
structure near the accreting binary system, marked by
the two small circles.
Arrows represent the flow direction (not scaled with velocity).
The preshock stream lines with 5 consecutive arrows represent the
limiting distance from the accretion line; stream lines closer
to the accretion line represent mass accreted to the binary
system behind the shock wave;
stream lines at larger distances from the accretion line
represent mass which escapes to infinity behind the shock.
The dense flow toward the accreting system along the
accretion line, represented by 3 close arrows, is
termed the accretion column. }
\end{figure}

I find now the appropriate value of $\gamma$ for the problem at hand.
The flow velocity from an AGB star is $v_s \simeq 10 \km \s^{-1}$.
The flow is mostly neutral and dusty.
The postshock temperature will be
$T_s \leq (3/16)\mu m_H v_s^2/k \simeq 2000 (v_s/10 \km \s^{-1})^2 \K$,
where $\mu m_H$ is the mean mass per particle, and $k$ is the
Boltzmann constant.
For such a postshock temperature a large fraction of the dust
and molecule are likely to survive the shock, and serve as
cooling agents for the gas.
The preshock density is given by
\begin{eqnarray}
\rho(a)= \frac {\dot M _a}{4 \pi a_0^2 v_s}
= 2 \times 10^{-16}  
\left( \frac {\vert \dot M_a \vert }{10^{-4} M_\odot \yr^{-1}} \right)
\left( \frac {v_s}{10 \km \s^{-1}} \right)^{-1}
\left( \frac {a_0}{100 \AU} \right)^{-2}
\g \cm^{-3} .
\end{eqnarray}
The post shock density will be a few times higher, becoming
even higher in the accretion column
(Pogorelov et al.\ 2000).
Using figure (11) of Woitke, Kr\"uger, \& Sedlmayr (1996), I find the
cooling time from $T \lesssim 8,000 \K$ and
density $\rho \gtrsim 10^{-16}\g \cm ^{-3}$ to be
$\tau_{\rm cool} \lesssim 1 \yr$.
For a density of $\rho \gtrsim 10^{-19} \g \cm^{-3}$ and
a temperature of $ T \lesssim 2,000 \K$, the cooling time is
still short, $\tau_{\rm cool} \lesssim 3 \yr$.
The typical flow time of most of the accreted gas is
\begin{eqnarray}
\tau_{\rm flow} \sim \frac {R_{\rm acc}}{v_s} = 5
\left( \frac{R_{\rm acc}}{10 \AU} \right)
\left( \frac{v_s}{10 \km \s^{-1}} \right)^{-1}
\yr.
\end{eqnarray}
I conclude that the postshock gas for the parameters relevant to
accretion from a wind of upper-AGB stars at orbital separations
where accretion disks are likely to be formed (see previous section)
is radiative. Namely, the gas is radiatively cooling quite efficiently,
such that the flow in the accretion column has a high Mach number,
and the effective adiabatic index is $\gamma \sim 1$.

The implication of the high Mach number and an efficient
radiative cooling is
that most of the mass is being accreted in a dense
and cold flow near the accretion line--the accretion column.
The case for an isothermal flow with very large Mach number,
i.e., pressure is negligible, has a simple solution
(e.g., Lyttleton 1972).
In this flow all stream lines hit the accretion line.
The total mass accretion rate per unit length on the accretion
line is constant $\dot m = \pi R_{\rm acc} \rho_0 v_0$,
where $\rho_0$ and $v_0$ are the density and velocity
of the unperturbed flow; here they will be taken as $v_0=v_s$,
and $\rho_0$ from equation (4).
The mass and momentum conservation equations for the flow
along the accretion line have infinite number of solutions for the
flow from the stagnation point outward, but only one solution
from the stagnation point toward the accreting body
(Lyttleton 1972; see his figure 1).
The equation for the velocity along the accretion line is
given by equation (6) in Lyttleton (1972).
The velocity of the flow $v(x)$, in units of $v_0$,
for the case of a stagnation point
at a distance of $x_s=R_{\rm acc}$ from the accreting body,
is presented in Figure 3; $x$ is the coordinate along the
accretion line measured from the accreting body, given
in the figure in units of $R_{\rm acc}$ (this is the
solution of equation [6] of Lyttleton, with his $\alpha=2$).
It is compared with the Keplerian velocity
along a circular orbit around the accreting body $v_K$.
\begin{figure}
\plotone{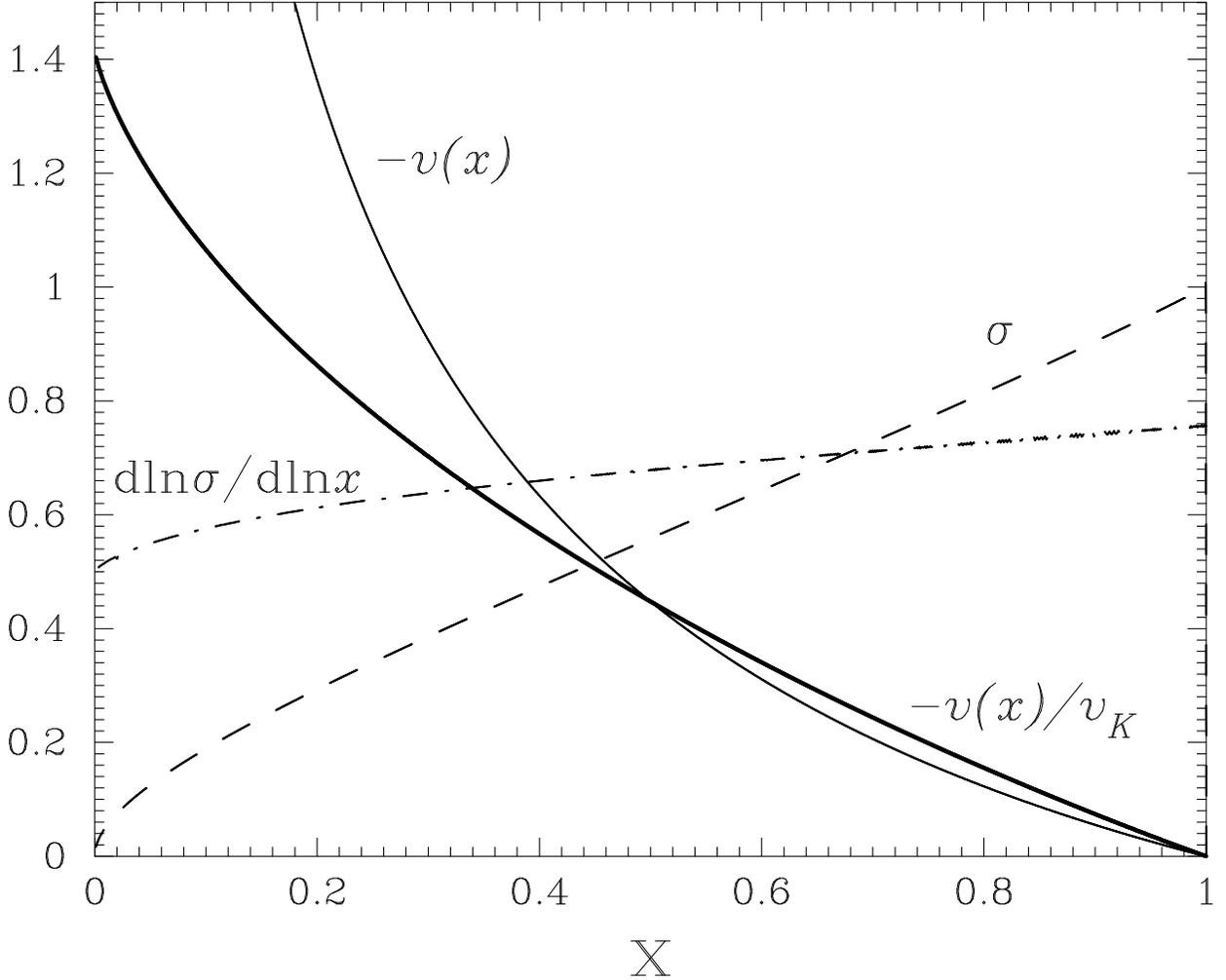}
\vskip 0.2 cm
\caption{The flow parameters along the accretion line behind
the center of mass of the accreting binary system, for
an isothermal accretion flow with large Mach number
(Lyttleton 1972).
Plotted are the magnitude of the velocity $-v(x)$ (thin
solid line), the ratio of the velocity to the Keplerian velocity in
a circular orbit around the binary system (thick solid line),
the density per unit length
in units of $2 \pi \rho_0 R_{\rm acc}^2$
(dashed line), and
$d \ln \sigma /d \ln x$ (dotted-dashed line).
$x$ is the distance from the center of mass given here in units of
$R_{\rm acc}=2 G M_{12}/v_0^2$, where $M_{12}=M_{b1}+M_{b2}$ is the total
mass of the accreting system and $v_0$ is the unperturbed velocity
at infinity. $x=R_{\rm acc}$ ($x=1$ in the figure) is the
stagnation point.  }
\end{figure}
The mass per unit length is given then by
$\sigma=\pi \rho_0 v_0 R_{\rm acc} (x_s-x)/\vert v(x) \vert$,
and it is plotted in figure 3 (dashed line)
in units of $2 \pi \rho_0 R_{\rm acc}^2$.

When accreting onto a binary system, this solution does not hold
any more at distances $x \lesssim a_1$, where $a_1$ is the
distance of the more massive star in the accreting binary system
from the center of mass of the binary system.

\section{THE SPECIFIC ANGULAR MOMENTUM OF THE ACCRETED MASS}
\subsection{General Considerations}
Each of the two stars, masses $M_{b1}$ and $M_{b2}$, accretes
mass mainly from the dense flow along the accretion
line, i.e., from the accretion column.
The accretion flow onto each star does not reach a steady state,
since each star periodically changes its distance from the
accretion column as it orbits the center of mass
(beside the case of exactly
perpendicular orbital planes; see subsection 3.2).
I consider now the accretion on star $M_{b1}$.
At closest approach of the star to the accretion column, its
orbital velocity is perpendicular to the velocity of
gas in the accretion column, and its relative
velocity to the gas is
\begin{eqnarray}
v_{r1} = [v_{K1}^2+v(a_1)^2]^{1/2} \sim 2^{1/2} v_{K}(a_1)=
\left[\frac {2 G M_{12}}{a_1} \right]^{1/2},
\end{eqnarray}
where $M_{12} \equiv M_{b1}+M_{b2}$, $v_{K1}$ is the orbital
velocity of the star around the
center of mass, $v_K$ is the Keplerian orbital velocity
around the binary system, $v(a_1)$ is the velocity of gas in the
accretion column relative to the center of mass of the
binary system, and $a_1=a_{12}M_{b2}/(M_{b1}+M_{b2})$
is the distance of the star from the center of mass of the binary system.
In the second equality I used the result presented in Figure 3,
concerning the inward flow in the accretion column, assuming that
$a_{12} \ll R_{\rm acc}$.
Assuming an accretion flow from the accretion column to the star
similar to a Bondi-Hoyle-Lyttleton flow, the accretion radius of the star
is $R_{\rm acc1} \simeq 2 GM_{b1}/v_{r1}^2$.
The ratio of the accretion radius of the star to its distance from
the center of mass is, using equation (6),
\begin{eqnarray}
\frac {R_{\rm acc1}} {a_1} \simeq \frac {M_{b1}}{M_{12}}.
\end{eqnarray}
If $M_{b2} =q M_{b1} \ll M_{b1}$, then little mass is accreted onto
the star $M_{b2}$. In addition, as will be seen later in this section,
the specific angular momentum of the mass accreted onto
the massive component $M_{b1}$ will be low, reducing the
probability of accretion disk formation (the limit of $q \ll 1$
gives the single accreting star case).

The exact conditions for the formation of accretion
disks around one or two stars must wait for full
3-dimensional numerical simulations.
In the present section I will estimate this value.
I will assume
\begin{eqnarray}
R_1 \ll w \lesssim a_{12} \ll R_{\rm acc} \ll a_0,
\end{eqnarray}
namely, the radius of the accreting star is much smaller than the
width of the accretion column at the binary location $w(x=a_{12})$,
which is smaller than orbital separation of the accreting binary
system, which is much smaller than the accretion radius of the
binary system (eq. 1), which is smaller than the orbital
separation with the mass losing star. For example, $30 \AU
\lesssim a_0 \lesssim 300 \AU$, $10 \AU \lesssim R_{\rm acc}
\lesssim 30 AU$, $10 R_\odot \simeq 0.05 \AU \lesssim a_{12}
\lesssim 1 AU$, with $w(x=a_{12})$ somewhat smaller, and $0.01
R_\odot \lesssim R_1 \lesssim 1 R_\odot$. As mentioned in section
1, the flow onto the binary system will not be exactly
axisymmetrical as depicted in figure 2, but rather the accreted
mass will posses some angular momentum. This excess angular
momentum causes the bending of the accretion column and shock
wave. However, for the large orbital separation $a_0$ assumed
here, this angular momentum and departure from axisymmetry of the
flow at large distance from the accreting binary system can be
neglected relative to the specific angular momentum of the
accreted mass within the binary system, as I now show. I start
with examining the situation for perpendicular orbital planes.

In the calculations to follow I assume that most accreted mass
reaches the binary system via the accretion column, such that the
characteristic width of the accretion column $w(x)$,
fulfills the condition $w(x=a_{12}) \lesssim a_{12}$.
When $w(x=a_{12}) \gtrsim a_{12}$ the calculations below do not hold
any more.
The binary system will be inside the accretion column,
the accretion process will be more complicated, and I expect that
the formation of an accretion disk will be less favorable.

\subsection{Perpendicular Orbital Planes}
In this case $\theta=90^\circ$ (see fig. 1).
Most of the accreted mass flows toward the center of mass
(again, the flow has a narrow accretion column, since
it is almost isothermal).
The gravitational attraction of the more massive star is stronger
on the accretion line, and most of the mass flows toward the massive companion.
I treat the more massive companion, a treatment holds for
equal mass components as well.

The angular momentum of mass element residing near the center
of mass relative to the center of the accreting star is
$j_{\rm cm}=\omega_{12} a_1^2$, where
$\omega_{12}=[G (M_{12})]^{1/2}/a_{12}^{3/2}$
is the angular velocity of the binary system.
For an accretion disk to form, this should be larger than
the specific angular momentum on the equator of the accreting star
$j_1=(G M_{b1} R_1)^{1/2}$.
The condition
\begin{eqnarray}
j_{\rm cm}=\omega_{12} a_1^2
 > j_1 = (G M_{b1} R_1)^{1/2}
\end{eqnarray}
reads
\begin{eqnarray}
\frac {a_{12}}{R_1} >
(1+q^{-1})^4 (1+q)^{-1} =
8
\left( \frac{M_{b2}}{0.5M_{12}}  \right)^{-4}
\left( \frac{M_{b1}}{0.5M_{12}}  \right) .
\end{eqnarray}
This condition can be met by a large fraction of binary systems
with mass ratio of $q \simeq 1$.
For $q=1$, a main sequence accretor, with $R_1 \simeq R_\odot$,
requires $a_{12} \gtrsim 10 R_\odot$,
and there are many binary systems with
$10 R_\odot \lesssim a_{12} \lesssim 1 \AU$.
However, for $q=0.5$, and $q=0.3$, the condition reads
$a_{12} > 54 {R_1}$, and $a_{12} > 271 {R_1}$, respectively.
For $q \ll 1$ the condition reads $a_{12} > q^{-4} {R_1}$.
Therefore, for $ q \lesssim 0.5$ only a small number of systems
with accreting main sequence stars are expected to form
accretion disks.
For accreting WDs, the mass ratio can be as small as
$ q \sim 0.1$.
However, such systems are rare and live for a short time, since
a WD mass is $\sim 0.5 M_\odot$, and a companion $\sim 10$ times
as massive, will evolve fast off the main sequence.
Over all, the formation of an accretion disk requires that the
mass ratio in most systems be $q \gtrsim 0.3$
(note that $q \leq 1$ by definition).

If the mass is accreted directly from the center of mass of the
accreting binary system, the accretion disk will be in the binary-orbital
plane, i.e., perpendicular to the orbital plane with the mass-losing star.
Some inclination is expected, though, since mass will
start flowing toward the accreting star before reaching the center of mass.
In any case, jets, if blown, will be in the orbital
plane of the triple-star system, or close to it.

\subsection{Parallel Orbital Planes}

In this case, each of the two stars in the accreting binary system
accretes mass in turn, when it reaches the accretion
column up-stream side.
The star in turn, say star $M_{b1}$, ``cleans'' the accretion column
up to a distance $R_{\rm acc1}$ from its location.
Namely, it cleans the region
$a_1-R_{\rm acc1} \lesssim  x({\rm clean}) \lesssim a_1+R_{\rm acc1}$,
where as before, $a_1$ is its distance from the
center of mass of the accreting binary system, and $R_{acc1}$ is its
accretion radius (eq. 7).
It will be convenient to work with the distance on the accretion
column measured from the location of the star $h \equiv x-a_1$,
and to approximate the density with
$\sigma(h)=\sigma(x=a_1)+\sigma^\prime h$, where
$\sigma^\prime \equiv d \sigma /dx$.

After one star leaves the vicinity of the accretion column,
some fraction of an orbit will elapse before the second
star starts to accrete mass.
If the stars accrete only when on the accretion line, the time
is exactly half a period.
The star, though, will influence the accretion line when approaching
and leaving the line as well.
Crudely, the accretion column has a time of $\sim 1/4 P_{12}$
to rebuild itself, where $P_{12} = 2 \pi / \omega_{12}$
is the orbital period of the binary system.
Under the assumption $a_{12} \ll R_{acc}$, the velocity
of the gas in the column density is of the order of
the orbital velocity of the binary system (fig. 3).
Therefore, during a time of $\gtrsim 0.2 P_{12}$ the gas
refills the accretion column, and I take the density per
unit length as given in figure 3.
The total accreted mass is then
\begin{eqnarray}
\Delta m = \int_{-R_{\rm acc1}}^{R_{\rm acc1}}
\sigma(h) d h = 2 R_{\rm acc1} \sigma(a_1).
\end{eqnarray}

The accreting star crosses the accretion line at a speed
$\omega _{12} a_1$, hence the specific angular momentum of mass
element at distance $h$ (along the accretion line) from the
accreting star is $\omega_{12} a_1 h$.
The total accreted angular momentum is
\begin{eqnarray}
\Delta J =
\int_{-R_{\rm acc1}}^{R_{\rm acc1}}
\sigma(h) \omega_{12} a_1 h d h = \omega_{12} a_1
\int_{-R_{\rm acc1}}^{R_{\rm acc1}}
[\sigma(a_1)+\sigma^\prime(a_1) h] h d h =
\frac{2}{3} \sigma^\prime(a_1) \omega_{12} a_1 R_{\rm acc1}^3.
\end{eqnarray}
From the last two equations the specific angular momentum
of the accreted mass is found to be
\begin{eqnarray}
j_{\rm par} = \frac {\Delta J}{\Delta m}=
\frac {1}{3}
\left( \frac {d \ln \sigma}{d \ln x} \right)_{x=a_1}
\left( \frac {R_{\rm acc1}}{a_1} \right)^2
\omega_{12} a_1^2.
\end{eqnarray}
The quantity $d \ln \sigma/d \ln x$ is plotted in Figure 3,
from which I approximate $(1/3) d \ln \sigma/d \ln x \simeq 0.2$
in the last equation.
I use  also equation (7) for $R_{\rm acc1}/a_1$ in equation (13).
The condition for the formation of an accretion disk
$j_{\rm par} > j_1$ (eq. 9), becomes
\begin{eqnarray}
\frac {a_{12}}{R_1} \gtrsim
25 (1+q^{-1})^4 (1+q)^{3}
= 3200
\left( \frac{M_{b2}}{0.5M_{12}}  \right)^{-4}
\left( \frac{M_{b1}}{0.5M_{12}}  \right)^{-3}.
\end{eqnarray}
This condition is impossible for main sequence stars to meet,
and almost impossible for WD accreting stars to meet.
Considering that the specific angular momentum of the accreted
mass is lower than that given by equation (13), as was shown
analytically (Davis and Pringle 1980) and numerically
(Livio et al.\ 1986; see the $\eta$ parameter in eq. [2]),
no accretion disk will be formed via accretion in the case
studied here, i.e, when the orbital planes are parallel.

\subsection{Inclined Orbital Planes}
From the two previous subsections we learned that the angular
momentum of the accreted mass is dominated by the component
parallel to the accretion line, namely, the accretion disk, if formed,
is perpendicular to the accretion line.
When the orbital planes are inclined by an angle $\theta$
(Fig. 1), the closest distance of the stars, say star $M_{b1}$,
to the accretion line is $a_1 \sin \theta$.
Most of the mass is accreted then, with a specific angular momentum
$j_{\rm inc} = \omega_{12} (a_1 \sin \theta)^2$.
The condition for the formation of an accretion disk $j_{\rm inc} > j_1$,
reads now (compared to eq. 10)
\begin{eqnarray}
\frac {a_{12}}{R_1} > 8
\left( \frac{M_{b2}}{0.5M_{12}}  \right)^{-4}
\left( \frac{M_{b1}}{0.5M_{12}}  \right)
\sin^{-4} \theta .
\end{eqnarray}

\section{DISCUSSION AND SUMMARY}
\subsection{Summary of Theoretical Calculations}

In the previous sections I calculated the specific angular momentum
of mass accreted onto a binary system, with the flow structure
as follows (section 2).
The binary system accretes from the dense wind of a mass losing
giant star, most relevant are AGB stars.
The flow is such that it is of the Bondi-Hoyle-Lyttleton type, i.e., the
orbital separation of the accreting binary system, $a_{12}$,
is much smaller than the Bondi-Hoyle accretion radius of the
system $R_{\rm acc}$ (eq. 1; fig. 1).
I also showed that for relevant parameters, the post-shock wind
cools fast relative to the flow time, leading to the formation of a
dense accretion flow behind the binary system: the accretion column.
I made the assumptions given in equation (8) regarding the
relations between the relevant length scales in the problem.
Typical length-scales are (see figures 1 and 2):
the radius of the accreting star (a WD or a main sequence star):
$0.01 R_\odot \lesssim R_1 \lesssim 1 R_\odot$;
the orbital separation of the stars in the accreting binary system:
$10 R_\odot \simeq 0.05 \AU \lesssim a_{12} \lesssim 1 AU$;
the Bondi-Hoyle accretion radius of the binary system (eq. 1):
$10 \AU \lesssim R_{\rm acc} \lesssim 30 AU$;
the orbital separation with the mass losing star:
$30 \AU  \lesssim a_0 \lesssim 300 \AU$.
I examine the formation of an accretion disk by demanding that
the specific angular momentum of the accreted mass be larger
than that of a test particle orbiting the equator of the accreting star.

The main results of these calculations can be summarized as follows.
\newline
(1) The accreted mass acquires angular momentum as a result of the
orbital motion of the accreting star around the center of mass of the accreting
binary system. When the orbital plane of the accreting binary system is
parallel to the orbital plane of the triple star system ($\theta=0$ in
fig. 1), the specific angular momentum is too small to form an accretion disk
for the assumed parameters, i.e., the condition for an accretion disk
formation is almost impossible to fulfill (eq. 14).
\newline
(2) When the orbital planes are perpendicular to each other, the
condition for disk formation is given by equation (10).
The accretion in this particular case is in a steady state,
because the distance of the stars from the accretion column does
not change.
The more massive star is expected to accrete most, or even all,
of the mass.
The condition can be easily met by many systems for which the mass ratio
is $q=M_{b2}/M_{b1} \gtrsim 0.3$.
\newline
(3) For an inclined orbit, $0^\circ < \theta < 90^\circ$,
the accretion occurs mainly when the accreting star is at its
closest approach to the accretion column.
The more massive star will accreted more mass,
but because of the periodic variation in distance from the
accretion column of both stars, the less massive star
accretes mass as well.
Assuming most of the mass is accreted indeed when the star
is close to the accretion column, the condition for disk formation
is given by equation (15).
This can be met by many systems if $\sin \theta$ is not too
small and $q \gtrsim 0.5$.
\newline
(4) For the calculations here assumption (8) was used.
This assumption is that
the radius of the accreting star is much smaller than
the width of the accretion column at the binary location,
which is smaller than orbital separation of the accreting
binary system, which is much smaller than the accretion
radius of the binary system (eq. 1), which is smaller than the
orbital separation with the mass losing star.
However, the basic physics can hold for a binary system very close
to the mass losing star. In that case, the mass in the accretion
column will have some angular momentum relative to the center
of mass of the accreting binary system, as single stars do
(the term $j_a$ in eq. 2).
The binary system may even have some tidal effect on the mass losing
giant star. The flow becomes very complicated in such a case.

With these and earlier results, the following conclusions can be drawn.
\newline
(5) For accreting binary systems which fulfill condition (15),
the constraint on disk formation becomes the mass accretion rate
as given by equation (3).
Because the mass accretion rate depends on total mass square, the
accretion rate is higher than in a single-star case.
Consider two equal mass main sequence
stars of $M_{b1}=M_{b2}=1 M_\odot$, and demanding that each star
accretes at a rate of  $> 10^{-7} M_\odot \yr^{-1}$.
For the mass loss rate and wind velocity as in equation (3), the
constraint from the mass accretion rate becomes
$a_0 \lesssim 400 \AU$.
For accreting WD stars, i.e., at least one of the stars in the binary
system is a white dwarf which accretes half of the mass, I take
$M_{b1}=M_{b2}=0.6 M_\odot$, and as in section 1 requires the
accretion rate to be $> 10^{-8} M_\odot \yr^{-1}$.
The constraint from the mass accretion rate becomes
$a_0 \lesssim 800 \AU$.
These numbers should be compared with the constraint on accreting single
stellar companion (first section) of $a_0 \lesssim 40 \AU$ and
$a_0 \lesssim 150 \AU$, for main sequence and WD single
companions, respectively.
In the later case the constraint is the accreted specific
angular momentum.
\newline
(6) The constraint on the mass accretion rate can be eased if we
consider the nature of the Bondi-Hoyle-Lyttleton accretion flow.
The flow along the accretion column is unstable,
with large variation in density, hence in the mass accretion
rate on short time scales (Cowie 1977; Soker 1991).
This implies that even when the average accretion rate is lower
than the critical value, on short time scales it can be higher,
leading possibly to sporadic jet formation.
\newline
(7) The formation of jets at large distance from the mass losing star
will lead to departure from axisymmetry, even when jets are blown
perpendicular to the triple-star orbital plane (Soker 2001).
Here there is another effect.
Because the accretion disk is almost perpendicular to the orbital plane,
i.e., its axis, hence the jets if are formed, is close to the
triple-star orbital plane and pointing from the the accreting binary
system to the mass-losing star.
This means that one jet expands toward lower density medium and expands almost
undisturbed.
The opposite jet, on the other hand, expands toward the mass losing star
and encounters dense medium. This jet can be deflected and/or slowed
down by the dense wind from the mass losing star.
\newline
(8) In addition to the accreting binary system which can reside at
a large orbital separation $a_0$, the mass losing giant star may
have a close companion. That closer companion will cause axisymmetrical mass
loss as well, but most likely with a different axis of symmetry.
\newline
(9) When $q \sim 1$, i.e., almost equal mass companions, both stars can
blow jets, each stars on its turn. This may further complicate
the structure of the nebula.
\newline
(10) The allowed large orbital separation for jets formation may
result in delayed jets (Soker 2001).
When the orbital separation to the mass losing star is
$a_0 \sim 200-500 \AU$, and the wind is slow
$v_0 \simeq 5-10 \km \s^{-1}$,
the time required for the wind to flow from the AGB
mass losing star to the accreting system is $\sim 100-500 \yr$.
This is a non-negligible fraction of the evolution time
during the super-wind phase.
The jets (or a CFW) may be blown after
the mass loss rate from the AGB star has been substantially reduced.
Namely, while the post-AGB star and its wind may show a post-AGB
age of 100-500 years, the jets may still be active!
\newline
(11) The triple-star scenario adds to the many routes through which binary
systems can shape PNs (Soker 2002).
This strengthens my earlier request not to use phrases like
`peculiar', `unique', `extraordinary', and
`unusual', in describing the kinematics and structure of PNs. All known
PNs with large departures from axi-symemtrical structures
can be fitted into the binary, including triple, stellar model
for the shaping of PNs.

\subsection{Observational Consequences}

The discussion in the preceding subsection shows that PNs descendant
from AGB stars which have two companions in a binary system that
blow jets, will most likely have a significant departure from
pure axisymmetrical structure. These systems may not have even a
mirror symmetry, because the two jets expands into different media.
The exact PN structure which result from jets blown by an accreting
binary system must be derived from 3D numerical simulations.
Here I only list some PNs which potentially were shaped by jets
blown by an accreting binary-system companion to their AGB progenitor.
Other types of systems may also lead to departure from both mirror
symmetry and from axisymmetry, e.g.,
stochastic mass loss from the mass losing star together with an
accreting companion which blows the jets,
or an accreting close single star companion, together with a wider
companion to the AGB star which sole role is to cause departure from
axisymmetry (Soker \& Hadar 2002).
Therefore, I don't expect all of the PNs listed below to have
triple star systems.
Some, though, may have triple star systems.
\newline
{\bf He 3-1375.} This PN has a dense ring, and small
lobes protrude from the nebula (Bobrowsky et al.\ 1998).
However, no single axis of symmetry, or point-symmetry, or
a plane of symmetry, can be defined to this nebula.
\newline
{\bf IC 2149 (PN G 166.1+10.4).} This PN has extremely
asymmetrical narrow structure (Balick 1987;
Vazquez et al.\ 2002). Vazquez et al.\ (2002) suggest that this is
the equatorial plane of the nebula. Another possibility is that
the narrow structure is composed of unequal jets.
\newline
{\bf M 1-59 (PN G 023.9-02.3).}
This bipolar PN has unequal-size lobes, both with departure from
axisymmetry (Manchado et al.\ 1996).
A close accreting single companion which blew two jets,
and a wider companion which sole role was to cause the departure
from axisymmetry, may also account for the structure, although this
model cannot by itself account for the unequal size of the two lobes.
\newline
{\bf NGC 6210 (PN G 043.1+37.7).} This is a `messy' PN,
with a general elliptical structure with unequal sides,
and blobs, filaments, and jet-like structures around it
(Balick 1987; Terzian \& Hajian 2000).
\newline
{\bf NGC 1514 (PN G 165.5-15.2)}
This PN has a general axisymmetrical structure, but with
large departure from exact axisymmetry revealed both in its
image (Balick 1987) and in its kinematics (Muthu \& Anandarao 2003).
\newline
{\bf NGC 6886(PN G 060.1-07.7).}
Two cylindrical-type lobes protrude from a spherical structure.
The two lobes are bent relative to the symmetry axis to the same side.
Bobrowsky et al.\ (2004) proposed that the lobes were formed by two
jets (or a collimated fast wind [CFW]) blown by a companion at orbital separation
of $\sim 30 AU$. The jets were bent by the ram pressure of the wind
from the mass losing AGB stellar progenitor of the PN.
This model by itself, however, cannot account for the observation
that the two lobes are unequal in size and intensity.

I thank an anonymous referee for helpful comments.
This research was supported in part by grants from the
Israel Science Foundation.



\end{document}